Please cite this article as:

Tao Xiong, Yukun Bao, Zhongyi Hu. "Multiple-output support vector regression with firefly algorithm for interval-valued stock price index forecasting". 2013, *Knowledge-based Systems*. 55, 2013:87-100.

# Research Highlights

➢ Extending the MSVR to the scenario of interval-valued time series forecasting.

➢ The parameters of MSVR are tuned using firefly algorithm (abbreviated to FA-MSVR).

➢ Assessing the forecasting ability of FA-MSVR on statistical and economic criteria.

➢ The experimental analysis is based on one- and multi-step-ahead forecasts.

➢ FA-MSVR is a promising method for interval forecasting of financial time series.

# Multiple-output support vector regression with a firefly algorithm for interval-valued stock price index forecasting

Tao Xiong, Yukun Bao[*], Zhongyi Hu

School of Management, Huazhong University of Science and Technology, Wuhan, P.R.China, 430074

## Abstract

Highly accurate interval forecasting of a stock price index is fundamental to successfully making a profit when making investment decisions, by providing a range of values rather than a point estimate. In this study, we investigate the possibility of forecasting an interval-valued stock price index series over short and long horizons using multi-output support vector regression (MSVR). Furthermore, this study proposes a firefly algorithm (FA)-based approach, built on the established MSVR, for determining the parameters of MSVR (abbreviated as FA-MSVR). Three globally traded broad market indices are used to compare the performance of the proposed FA-MSVR method with selected counterparts. The quantitative and comprehensive assessments are performed on the basis of statistical criteria, economic criteria, and computational cost. In terms of statistical criteria, we compare the out-of-sample forecasting using goodness-of-forecast measures and testing approaches. In terms of economic criteria, we assess the relative forecast performance with a simple trading strategy. The results obtained in this study indicate that the proposed FA-MSVR method is a promising alternative for forecasting interval-valued financial time series.

**Keywords:** Stock price forecasting; Interval-valued data; Multiple-output support

[*] Corresponding author: Tel: +86-27-87558579; fax: +86-27-87556437.
Email: yukunbao@hust.edu.cn or y.bao@ieee.org

vector regression; Firefly algorithm; Trading strategy.

# 1. Introduction

Forecasting stock prices is a fascinating issue in financial market research. Accurately forecasting stock prices, which forms the basis for decision making regarding financial investments, is most likely the greatest challenge for the capital investment industry and is thus of great interest to academic researchers and practitioners.

According to an extensive literature investigation, it is not difficult to find a wide variety of methodologies and techniques that have been used for stock price forecasting with various degrees of success, such as the Box–Jenkins method [1], general autoregressive conditional heteroskedasticity [2], stochastic volatility model [3], fuzzy logic approach [4], grey-based approaches [5], wavelet transforms and adaptive models [6], neural networks [7], support vector regression [8, 9], hybrid models [10, 11], and decision support systems [12]. However, it is important to note that the studies above considered point forecasting rather than interval forecasting.

Interval forecasts of stock prices have the advantage of taking into account variability and/or uncertainty, reducing the amount of random variation relative to that found in classic single-valued stock price time series (e.g., stock closing price). As Hu and He [13] noted, the interval forecasts of stock price are superior to the traditional point forecasts in terms of the overall lower mean error and higher average accuracy ratio. Moreover, intervals of stock prices have been widely used in the construction of a variety of technical trading rules [14]. To date, there has been a great deal of research focused on exploring the underlying dynamics of interval-valued stock prices and developing suitable models for forecasting them [13, 15-18]. For example, Maia, et al. [15] proposed a hybrid methodology that combines the ARIMA and ANN models for interval-valued stock price forecasting. Cheung, et al. [18] found evidence of cointegration between the daily log highs and log lows of several major stock indices and further forecasted the daily log highs and log lows using a vector error correction model (VECM). The reader is referred to Arroyo, et al. [16] for a recent survey of the present methodologies and techniques employed for interval-valued stock price forecasting. It should be noted that the interval-valued data in this study do not come from noise assumptions as in [19], but rather from the expression of variation or the aggregation of huge databases into a reduced number of groups as in [13, 15-18].

Our study focuses on extending the multi-output support vector regression (MSVR) to adapt to the scenario of interval forecasting of a stock price index. As a well-known intelligent algorithm, support vector regression (SVR) [20] has attracted particular attention from both practitioners and academics for use in time series forecasting during the last decade. SVR algorithms have been found to be a viable contender among various time-series models [21, 22] and have been successfully applied to different areas [23]. Despite the promising SVR demonstrated in [21-24], the applications of SVR in interval-valued time series (ITS) have not been widely explored. This is because the standard formulation of SVR can only be used as a univariate modeling technique for ITS forecasting due to its inherent single-output structure. The univariate technique fits and forecasts the interval bounds of ITS independently, without considering the possible interrelations that are present among them, which has been criticized in [16]. To generalize the SVR from a regression estimation to multi-dimensional problems, Pérez-Cruz, et al. [25] proposed a multi-dimensional SVR that uses a cost function with a hyperspherical intensive zone, which is capable of obtaining better predictions than using an SVR model independently for each dimension. Recently, Tuia, et al. [26] proposed a multi-output SVR model (MSVR), based on the previous contribution in [27], for simultaneously estimating different biophysical parameters from remote sensing images. In the work of [25, 27], the MSVRs are established and justified in a variety of disciplines [26-28]. Although past studies have clarified the capability of the MSVR, there have been very few, if any, efforts to evaluate the performance of the MSVR for time series forecasting, particularly interval-valued time series forecasting. As such, we set out to investigate the possibility of forecasting the lower and upper bounds of stock index series simultaneously by making use of an MSVR. In this model, the inputs of the MSVR are the lagged intervals, while the two outputs of the MSVR correspond to the forecasts of the bounds.

Parameter selection for the MSVR is another issue addressed in this study. The generalization ability of the MSVR depends on adequately setting parameters, such as the penalty coefficient $C$ and kernel parameters. Therefore, the selection of optimal parameters is crucial to obtain good performance in handling forecasting tasks with MSVR. To date, a large number of evolutionary algorithms, such as the genetic algorithm (GA) and particle swarm optimization (PSO), have been employed to optimize the parameters of SVR. The firefly algorithm (FA), a novel swarm-based intelligent metaheuristic, was recently introduced by Yang [29]. The FA mimics the social behavior of fireflies, which move and communicate with each other based on characteristics such as the brightness, frequency and time period of their flashing. The superiority of the FA against the GA and PSO in existing studies [29-31] motivates us to use the FA for selecting parameters for the MSVR. By doing so, this study proposes a FA-based approach to appropriately determine the parameters of MSVR for ITS forecasting (abbreviated as FA-MSVR). For comparison purposes, a univariate technique (fitting the interval bounds independently), standard SVR, and three well-established interval-valued forecasting methods (fitting the interval bounds simultaneously), namely Holt's exponential smoothing method for intervals (Holt$^I$) [17], the vector error correction model (VECM) [18], and the interval multilayer perceptron (iMLP) [32], are selected as benchmarks. Three globally traded broad market indices, the S&P 500, FTSE 100, and Nikkei 225, are chosen as experimental datasets.

To examine the performance of the proposed FA-MSVR method for interval forecasts of a stock price index, we analyze the out-of-sample one- and multi-step-ahead forecasts from the FA-MSVR and selected benchmarks in two ways. First, we examine whether the out-of-sample forecasts generated by the FA-MSVR are more accurate than and preferable to those generated by the benchmark methods for an interval-valued stock index series, employing statistical criteria such as the goodness of forecast measure (e.g., the interval's average relative variance) and the accuracy compared to competing forecasts (e.g., the analysis of variance test and Tukey's HSD test). Second, we analyze whether the FA-MSVR is superior to the selected benchmarks in practice, assessing its relative forecast performance with economic criteria. We use the forecasts of lower and upper bounds from the different methods in a simple trading strategy and compare the returns to determine whether the FA-MSVR is a useful forecasting approach for an investor.

In summary, our contributions could be outlined as follows. First, we extend the MSVR in a novel manner to adapt to the scenario of interval-valued time series forecasting. Second, the possibility of forecasting the lower and upper bounds of interval-valued stock index series simultaneously by the established MSVR is examined. Third, to address the determination of parameters for the MSVR, the parameters of MSVR are tuned using a recently proposed FA. Finally, not only statistical accuracy but also economic criteria are used to assess the practicability of the FA-MSVR for interval-valued stock index forecasting.

This paper is structured as follows. In Section 2, we provide a brief introduction to the MSVR and illustrate the data representation of an interval-valued stock index series analysis. Afterwards, the proposed FA-MSVR method is discussed in detail in Section 3. Section 4 details the research design of the data descriptions, statistical and economic criteria, input selection, and implementation of the methodologies. Following that, in Section 5, the experimental results are discussed. Section 6 concludes this work.

## 2. MSVR with an interval-valued stock index series

This section presents the overall formulation process of the MSVR for interval-valued stock index series forecasting. First, the data representation of interval-valued stock index series is illustrated. Then, the MSVR for the obtained ITS is formulated in detail.

### 2.1 Construction of an interval-valued stock index series

An interval-valued variable, $[X]$, is a variable defined for all of the elements $i$ of a set $E$, where $[X_i] = \left\{ \left[ X_i^L, X_i^U \right]^T : X_i^L, X_i^U \in \mathbb{R}, X_i^L \leq X_i^U \right\}$, $\forall i \in E$. Table 1 shows the daily interval values of the S&P 500 index.

**<Insert Table 1 here>**

The particular value of $[X]$ for the $i$th element can be denoted either by the interval's lower and upper bounds, $[X_i] = \left[ X_i^L, X_i^U \right]^T$, or the center (mid-point) and radius (half-range), $[X_i] = \left[ X_i^C, X_i^R \right]^T$, where $X_i^C = \left( X_i^L + X_i^U \right) \big/ 2$ and $X_i^R = \left( X_i^U - X_i^L \right) \big/ 2$. Fig. 1 illustrates the structure of an interval.

**<Insert Fig. 1 here>**

An interval-valued time series (ITS) is a chronological sequence of interval-valued variables. The value of a variable at each instant in time $t \, (t = 1, \ldots, n)$ is expressed as a two-dimensional vector $\left[ X_t^L, X_t^U \right]^T$ with the elements in $\mathbb{R}$ representing the lower bound $X_t^L$ and upper bound $X_t^U$, with $X_t^L \leq X_t^U$. Thus, an ITS is $[X_t] = \left[ X_t^L, X_t^U \right]^T$ for $t = 1, \ldots, n$, where $n$ denotes the number of intervals of the time series (sample size). Fig. 2 illustrates the stock market in which a daily interval-valued S&P 500 index series arises. Fig. 2(a) illustrates a 10-minute S&P 500 index from Dec. 11, 2012, to Dec. 27, 2012. Fig. 2(b) depicts the corresponding daily S&P 500 index intervals.

**<Insert Fig. 2 here>**

## 2.2 MSVR for interval-valued stock index forecasting

The MSVR, proposed by Pérez-Cruz, et al. [25] to solve the problem of regression estimation for multiple variables, is a generalization of the standard SVR. Detailed discussions of the MSVR can be found in [25-27], but a brief introduction about the formulation of the MSVR for interval-valued time series forecasting is provided here.

Begin with an interval-valued stock index series $\left[ X_t \right] = \left[ X_t^L, X_t^U \right]^T$ for $t = 1, \ldots, n$ as shown in Fig. 2(b), where $n$ denotes the number of intervals of the time series and $X_t \in \mathbb{R}^2$ represents the $t$th interval. Interval-valued stock index series modeling and prediction is defined as finding the mapping between an input vector $\mathbf{x} = \left[ X_i^L, X_i^U, X_{i-1}^L, X_{i-1}^U, \ldots, X_{i-d+1}^L, X_{i-d+1}^U \right]^T \in \mathbb{R}^{2d}$ and an output vector $\mathbf{y} = \left[ X_{i+1}^L, X_{i+1}^U \right]^T \in \mathbb{R}^2$ from a given independent, identically distributed sample, i.e., $\left\{ \left( \mathbf{x}_i, \mathbf{y}_i \right) \right\}_{i=d}^n$. As such, we use an MSVR model with $2d$ inputs (lagged intervals at $i, i-1, \ldots, i-d+1$) and two outputs, with each output corresponding to the forecast of the bounds, $X_{i+1}^L$ and $X_{i+1}^U$.

The MSVR solves the problem above by finding the regressors $\mathbf{w}^j$ and $b^j$ ($j$=1, 2) for every output that minimizes:

$$L_p \left( \mathbf{W}, \ \mathbf{b} \right) = \frac{1}{2} \sum_{j=1}^2 \left\| \mathbf{w}^j \right\|^2 + C \sum_{i=1}^n L(u_i) \tag{1}$$

where $u_i = \|\mathbf{e}_i\| = \sqrt{(\mathbf{e}_i^T \mathbf{e}_i)}$,

$\mathbf{e}_i^T = \mathbf{y}_i^T - \varphi(\mathbf{x}_i)\mathbf{W} - \mathbf{b}^T$

$\mathbf{W} = \left[\mathbf{w}^1, \mathbf{w}^2\right]$,

$\mathbf{b} = \left[b^1, b^2\right]^T$

$\varphi(\cdot)$ is a nonlinear transformation of the feature space, which is usually a higher-dimensional space, and $C$ is a hyper parameter that determines the trade-off between the regularization and error reduction terms. $L(u)$ is a quadratic epsilon-insensitive cost function defined in Eq. (2), which is a differentiable form of the Vapnik $\varepsilon$ insensitive loss function.

$$L(u) = \begin{cases} 0 & u < \varepsilon \\ u^2 - 2u\varepsilon + \varepsilon^2 & u \geq \varepsilon \end{cases} \qquad (2)$$

In Eq. (2), when $\varepsilon$ is nonzero, the function will take all outputs into account when constructing each individual regressor and will obtain more robust predications and will then yield a single support vector set for all dimensions. It should be noted that the proposed problem cannot be resolved in a straightforward manner; thus, an iterative reweighted least squares (IRWLS) procedure based on a quasi-Newton approach to obtain the desired solution was proposed by [27]. By introducing a first-order Taylor expansion of the cost function $L(u)$, the objective of Eq. (1) will be approximated by the following equation:

$$L'_p(\mathbf{W}, \ \mathbf{b}) = \frac{1}{2}\sum_{j=1}^{2}\left\|\mathbf{w}^j\right\|^2 + \frac{1}{2}\sum_{i=1}^{n}a_i u_i^2 + CT, \quad a_i = \begin{cases} 0 & u_i^k < \varepsilon \\ \dfrac{2C(u_i^k - \varepsilon)}{u_i^k} & u_i^k \geq \varepsilon \end{cases} \qquad (3)$$

where $CT$ is a constant term that does not depend on $\mathbf{W}$ and $\mathbf{b}$, and the superscript $k$ denotes the $k$th iteration.

To optimize Eq. (3), an IRWLS procedure is constructed that linearly searches the next step solution along the descending direction based on the previous solution [27]. According to the Representer Theorem [33], the best solution for minimizing Eq. (3) in the feature space can be expressed as $\mathbf{w}^j = \sum_i \beta_i^j \phi(\mathbf{x}_i)$, so the target of the MSVR becomes finding the best $\boldsymbol{\beta}$ and $\mathbf{b}$. The IRWLS of MSVR can be summarized in the following steps [27, 28]:

*Step 1*: Initialization: Set $k = 0$, $\boldsymbol{\beta}^k = 0$, and $\mathbf{b}^k = 0$ and calculate $u_i^k$ and $a_i$;

*Step 2*: Compute the solutions $\boldsymbol{\beta}^s$ and $\mathbf{b}^s$ according to the following equation:

$$\begin{bmatrix} \mathbf{K} + \mathbf{D}_a^{-1} & 1 \\ \mathbf{a}^T \mathbf{K}^a & 1^T \mathbf{a} \end{bmatrix} \begin{bmatrix} \beta^j \\ b^j \end{bmatrix} = \begin{bmatrix} \mathbf{y}^j \\ \mathbf{a}^T \mathbf{y}^j \end{bmatrix}, \quad j = 1, 2 \tag{4}$$

where $\mathbf{a} = [a_1, \ldots, a_n]^T$, $(\mathbf{D}_a)_{ij} = a_i \delta(i - j)$, and $\mathbf{K}$ is the kernel matrix. Define the corresponding descending direction $\mathbf{p}^k = \begin{bmatrix} \mathbf{w}^s - \mathbf{w}^k \\ (\mathbf{b}^s - \mathbf{b}^k)^T \end{bmatrix}$.

*Step 3*: Use a backtracking algorithm to compute $\boldsymbol{\beta}^{k+1}$ and $\mathbf{b}^{k+1}$ and further obtain $u_i^{k+1}$ and $a_i$. Return to step 2 until the function converges.

The convergence proof of the above algorithm is given in [27]. Because $u_i^k$ and $a_i$ are computed by means of every dimension of $\mathbf{y}$, each individual regressor contains information about all outputs, which can improve the prediction performance [28].

Here, we selected the radial basis function (RBF) as the kernel function. To determine the parameters, namely $C$, $\varepsilon$, and $\sigma$ (in the case of RBF as the kernel function), of the MSVR, we develop a FA-based approach for parameter determination of the MSVR (termed FA-MSVR), which is detailed in the following section.

## 3. The proposed FA-MSVR method

In this section, the overall formulation process of the FA-MSVR for ITS forecasting is presented. First, the FA is briefly introduced. Then, the proposed FA-MSVR method is formulated and the procedure is presented in detail.

### 3.1 Firefly algorithm

The firefly algorithm (FA), first introduced by Yang [29], is a swarm-based intelligent metaheuristic. In the FA, each firefly is assumed to be attracted to all other fireflies regardless of their sex, and their attractiveness is proportional to their brightness. The brightness of a firefly is determined by the fitness function. The movement of a firefly $x_i^t$ attracted by another, more attractive firefly $x_j^t$ can be formulated as:

$$x_i^{t+1} = x_i^t + \beta_0 e^{-\gamma r_{ij}^2} \left( x_j^t - x_i^t \right) + \alpha \left( rand - \frac{1}{2} \right) \tag{5}$$

where the second term is the attraction of firefly $x_j^t$ to firefly $x_i^t$, the third term is the randomization of the movement, $\gamma$ is an absorption coefficient, $r_{ij} = \left\| x_j^t - x_i^t \right\|$ is the Cartesian distance between the two fireflies $x_i^t$ and $x_j^t$, $\beta_0$ is the attractiveness at $r_{ij} = 0, \alpha$ is a randomization parameter, and *rand* is a random number generator uniformly distributed in [0,1].

Here, we briefly analyze the absorption coefficient $\gamma$. For a detailed illustration of the other parameters mentioned above, please refer to [29, 31, 34]. From Eq. (5), it is clear that there exist two important limiting cases when $\gamma \to 0$ and $\gamma \to \infty$. For $\gamma \to 0$, the attractiveness is constant, so that a firefly can be seen by all of the other fireflies. On the contrary, for $\gamma \to \infty$, the attractiveness is almost zero from the perspective of other fireflies. This means that the fireflies fly randomly through a very foggy region, which corresponds to a random search algorithm. As the FA is usually somewhere in between these two extremes, it is possible to fine-tune these parameters so that the FA can outperform both the PSO and a random search [34]. To further assess the performance of the proposed FA-MSVR for interval-valued stock price index forecasting, we compare the experimental results of the FA-MSVR with those produced by other evolutionary algorithms, such as particle swarm optimization (PSO) and genetic algorithm (GA). The results and discussions are given in Appendix A to save space.

## 3.2 The FA for selecting parameters for the MSVR

This study developed an FA approach, termed FA-MSVR, for determining the parameters of the MSVR. Fig. 3 lists the pseudocode algorithm for the FA-MSVR.

**\<Insert Fig. 3 here\>**

As mentioned in Section 2.2, the RBF is selected as the kernel function. Thus, the three decision variables, designated $C$, $\sigma$, and $\varepsilon$, are required. As Hsu, et al. [35] suggested, an exponentially increasing sequence is a practical method to identify optimal parameters. Hence, the search space of these parameters is defined as an exponentially growing space: $\log_2 C \in [-6,6]$, $\log_2 \sigma \in [-6,6]$, and $\log_2 \varepsilon \in [-6,6]$.

The fitness of each firefly is the prediction performance of the MSVR for an interval-valued stock price index series in terms of the interval average relative variance (ARV[I]), a classical error measurement adapted in [17] for ITS problems.

Lower ARV$^I$ values lead to better forecasts [17]. The definition of ARV$^I$ is shown as follows:

$$\text{ARV}^I = \frac{\sum_{j=1}^{n}\left(X_{j+1}^U - \hat{X}_{j+1}^U\right)^2 + \sum_{j=1}^{n}\left(X_{j+1}^L - X_{j+1}^L\right)^2}{\sum_{j=1}^{n}\left(X_{j+1}^U - \overline{X}^U\right)^2 + \sum_{j=1}^{n}\left(X_{j+1}^L - \overline{X}^L\right)^2} \quad (6)$$

where $n$ denotes the number of fitted intervals, $\left[X_t^L, X_t^U\right]$ is the $t$th true interval, $\left[\hat{X}_t^L, X_t^U\right]$ is the $t$th fitted interval, $\left[\overline{X}^L, \overline{X}^U\right]$ is the sample average interval, and $\overline{X}^L$ and $\overline{X}^U$ are the lower and upper bound averages, respectively.

## 4. Research design

This section provides details about the research design of the data description, statistical and economic criteria, input selection, and implementation of the methodology. Further experimental results and discussion are reported in the next section.

### 4.1 Data descriptions

Three globally traded broad market indices, namely S&P 500 for the US, FTSE 100 for the UK, and Nikkei 225 for Japan, are chosen as experimental datasets[1]. The sample data are daily interval-valued data of three indices with different periods and sample sizes, as shown in Table 2. The intervals are obtained for a daily range of a selected stock index; the lowest and highest values for the day are calculated to define the movement in the market for that day. The data are expressed on a log scale. For each stock index, the first two-thirds of the observations are used as an estimation sample, while the remainder are reserved as the hold-out sample. Each of the examined methods was implemented (or trained) on the estimation sample, and forecasts were produced for the entire hold-out sample. The forecasts were then compared to the hold-out sample to evaluate the out-of-sample performance of each method. A five-fold cross validation was used in the training phase to avoid over-fitting. For the purpose of a description of the dataset, the interval-valued S&P 500 index is used as an example (see Fig. 4). The interval-valued S&P 500 index contains 523 observations from July 19, 2010, to August 10, 2012. The first 349 observations from July 19, 2010, to December 1, 2011 are used as the estimation sample, and the last 174 observations from December 2, 2011 to August 10, 2012 are reserved as the hold-out sample.

---

[1] Free data are available from the Yahoo Finance website (http://finance.yahoo.com/)

<center>**<Insert Table 2 here>**</center>

<center>**<Insert Fig. 4 here>**</center>

We employ one-step- and multi-step-ahead forecasts because it is useful to compare the selected methods at higher horizons. For multi-step-ahead predictions, the iterated strategy is implemented in our study. This strategy constructs a prediction model by means of minimizing the squares of the in-sample one-step-ahead residuals and then uses the predicted value as an input for the same model when forecasting the subsequent point, continuing in this manner until reaching the horizon.

## 4.2 Statistical and economic criteria

To assess the predictive ability of the different methods, we compare the out-of-sample forecasts using two different approaches because it is generally impossible to specify a forecast evaluation criterion that is universally acceptable.

First, we examine the forecast accuracy of all of the estimated methods by calculating the ARV$^I$, which is defined in Eq. (6). Second, we employ various tests of hypotheses; we use the analysis of variance (ANOVA) test to determine if a statistically significant difference exists among the estimated methods in out-of-sample forecasts. To further identify the significant differences between any two methods, the Tukey's HSD test [36] is used to compare all pair-wise differences simultaneously. Note that the Tukey's HSD test is a post-hoc test, meaning that Tukey's HSD test should not be performed unless the results of the ANOVA test are positive.

As discussed in Section 1, the highs and lows of assets have been widely used in the construction of a variety of technical trading rules in stock markets. Knowledge of these qualities is also extremely relevant to analyzing extreme price movements as well as volatility [37]. Thus, to examine the predictive ability of the different methods in practice, we assess the relative forecast performance with economic criteria. To do so, we use the forecasts of lower and upper bounds from the different methods in a simple trading strategy and compare the returns, i.e., the average annualized returns and percentage of trades resulting in positive returns from any trade, to determine which methods are superior for an investor. Let $X_t^O$ and $X_t^C$ be the opening and closing values, respectively, of a stock index on day $t$ and let $\hat{X}_{t+h}^L$ and $\hat{X}_{t+h}^U$ be the predicted low and high of the stock index for day $t+h$, formed after the market closes on day $t$.

The trading rule works as follows [14, 37]:

*Step 1*: On a given day $t$, a 'buy' signal for the stock index is generated if

$$\hat{X}_{t+h}^{U} - X_t^O > X_t^O - X_{t+h}^L.$$

*Step 2*: Buy the stock index on day $t + k - 1$ using the closing price of that day if the 'buy' signal is observed for $k$ consecutive days beginning with day $t$.

*Step 3*: On another day $s$ subsequent to buying the asset, generate a 'sell' signal if

$$\hat{X}_{s+h}^{U} - X_s^O < X_s^O - X_{s+h}^L.$$

*Step 4*: Sell the asset using the closing price $\hat{X}_{s+k-1}^C$ of day $s + k - 1$ as the execution price if a 'sell' signal has been observed for $k$ consecutive trading days beginning with day $s$. Otherwise, hold the asset.

It is worth noting that the prediction horizon $h$ and the observed consecutive trading days $k$ should be predefined; they do not change in the sequence of steps. For the purpose of illustration, a simple trading strategy with $h = 2$ and $k = 5$ is raised as an example. Suppose that according to the condition of *Step 1*, a 'buy' signal emerges on day $t = 6$. Now, if the 'buy' signal is observed also on days 7, 8, 9, and 10 in addition to day 6 (i.e., observed for $k = 5$ consecutive trading days), then by *Step 2*, the investor will buy the asset on day 10 ($t + k - 1 = 10$). Otherwise, the current state is preserved. Now, suppose that according to the condition of *Step 3*, a 'sell' signal emerges on day $s = 12$. If the 'sell' signal is also observed on days 13, 14, 15, and 16, then by *Step 4*, the investor will sell the asset on day 15. Otherwise, the current state is preserved.

## 4.3 Input selection

The filter method, which selects a set of inputs by optimizing a criterion over different combinations of inputs by means of a search algorithm, is employed for selecting the inputs of the FA-MSVR in this study. The filter method requires setting two elements: the relevance criteria, i.e., a statistic that estimates the quality of the selected variables, and the search algorithm, which describes the policy to explore the input space. Specifically, concerning the relevance criteria, the 2-fold cross-validation of a 1-NN approximator [38], as shown in the following pseudo-code in Fig. 5, is adopted. With respect to the search algorithm, a forward-backward selection method that offers the flexibility to reconsider previously discarded input variables and to discard previously selected input variables is used. The maximum embedding order, $d$, is set to 12 by the rule of thumb.

**<Insert Fig. 5 here>**

## 4.4 Implementation of methodologies

As discussed in Section 1, a univariate technique (the standard SVR [20]) and three interval-valued forecasting methods (Holt[I] [17], VECM [18], and iMLP [32]) are selected as benchmarks. The essential formulations of these selected methods have been presented in many papers, so they will not be repeated here to keep this paper concise. For a detailed introduction to these methods, please refer to [17, 18, 20, 32].

Taking into account the number of interval-valued stock index series (that is, 3), it is necessary to create estimates for 3 models using FA-MSVR, Holt[I], VECM, and iMLP, one for each index, and 6 models undertaking SVR, as the SVR is applied independently to forecast the lower and upper bounds of a given interval stock index series. The implementations of these methods are described in detail below.

The proposed FA-MSVR model is implemented in a Matlab computing environment. Specifically, the MSVR with two outputs is implemented using the Matlab program[2] provided by Pérez-Cruz, et al. [25]. Based on the pseudocode for the FA presented in Yang [29], the FA is implemented in Matlab. Selecting the parameters (the population size $I$, the absorption coefficient $\gamma$, the attractiveness $\beta_0$ at $r = 0$, and the randomization parameter $\alpha$) for the FA is yet another challenging model selection task. Fortunately, several empirical and theoretical studies have been performed about the parameters of the FA from which valuable information can be obtained [29, 31, 34]. In this study, the population size $I$ is determined through preliminary simulation. The effects of population size on prediction accuracy are shown in Table 3. Looking at Table 3, it is clear that, with the increase of population size, the prediction accuracy varies with small magnitude. That is to say, the population size does not affect the searching quality of FA-MSVR too much. Thus, it is recommended to adopt a population size of 20 for problems with small and medium size as like in this case. The rest of parameters are selected according to the recommendations in [29, 31, 34]. As such $I = 20$, $\gamma = 1$, $\beta_0 = 1$, and $\alpha \in [0,1]$ are used herein.

**<Insert Table 3 here>**

LibSVM (version 2.86) [39] is employed for the standard SVR in this study. The RBF is selected as the kernel function in the SVR. To determine the parameters of the SVR, a straightforward grid search is employed.

The Holt[1] is adopted here for interval time series, as is done in [17]. The smoothing parameter matrices with elements constrained to the range (0, 1) can be estimated by minimizing the interval sum of the squared one-step-ahead forecast errors. The solution of this optimization problem can be obtained using the limited memory BFGS method for bound constrained optimization (L-BFGS-B), which has been implemented in the R software package 'optimx'[3].

---

[2] Source code is available at http://www.uv.es/gcamps/code/msvr.htm

For the VECM estimation, we first conduct the preliminary analyses using the example of a daily interval-valued S&P 500 index series as shown in Fig. 4. The augmented Dickey-Fuller (ADF) results at 0.05 levels of significance confirm that these series are non-stationary in level but stationary in first differences. These results call for a formal test of cointegration between $X_i^L$ and $X_i^U$. Thus, the Johansen test is used to investigate if there are any cointegrated relations between the variables. The Bayesian criterion is used to select the lag parameter $p$. According to both maximum eigenvalue and trace statistics, the null hypothesis of no cointegration is rejected at 0.05 levels of significance (see Table 4). Furthermore, there is no evidence that more than one cointergrating vector exists. We thus set the dimension of the cointegration space to 1, that is, the lower and upper bounds series of the daily S&P 500 index from July 19, 2010, to August 10, 2012 are considered to be CI(1,1). The identical preliminary analyses mentioned above are conducted for the FTSE 100 and Nikkei 225 index as well. These results suggest that the lower and upper bounds of each daily stock index considered are integrated. Thus, a VECM is the natural empirical construct to examine their long-run and short-run interactions. As such, the VECM for the daily interval-valued stock index is implemented using Eviews.

---

[3] R package 'optimx' is available at http://ftp.ctex.org/mirrors/CRAN/

<center>**<Insert Table 4 here>**</center>

The iMLP is adopted here for the interval time series, as is done in [32]. Based on the structure of the iMLP presented by Roque, et al. [32], the iMLP is implemented in a Matlab computing environment. To minimize the cost function formulated in [32], the BFGS quasi-Newton method and back propagation procedure have been applied. An iMLP with 15 neurons in hidden layer has been trained with an estimation sample.

## 5. Experimental results

This section focuses on the out-of-sample forecasting ability of the examined methods in terms of statistical accuracy (Section 5.1) and economic criteria (Section 5.2). The analysis is based on one- and multi-step-ahead forecasts, considering short and long forecast horizons $\left( h = 1, 3, 5 \right)$.

### 5.1 Statistical assessment of the out-of-sample forecasts

This section demonstrates the usefulness of the methods through statistical evaluations in experiments using three interval-valued stock index series. The main objective of the experiments is to compare the performances of the FA-MSVR proposed in this paper with those of four selected benchmarks in terms of accuracy measure (the ARV[I]) and equality of accuracy of competing forecast tests (the ANOVA and HSD tests), to forecast an interval-valued time series in a stock market.

For each stock index, the data are first split into the estimation sample and the hold-out sample. Then, the input selection and model selection for the estimation sample are determined using the aforementioned filter method and a five-fold cross-validation technique, respectively. Next, the attained models are tested on hold-out samples. Afterward, the $ARV^I$ are computed for each prediction horizon. We repeat the previous modeling process 50 times, yielding 50 $ARV^I$ for each method and prediction horizon. Upon the termination of this loop, the performances of the examined methods at each prediction horizon are judged in terms of the mean of the $ARV^I$ of the 50 replications for hold-out samples. The experiments are detailed below. Figs. 6-8 show the comparisons of the performance of the different methods in all prediction horizons across the three indices.

**<Insert Figs. 6-8 here>**

As per the results presented, one can deduce the following observations. Overall, the top three methods across the three indices and three prediction horizons were the FA-MSVR, and then the VECM and the iMLP almost tied. It is clear that the proposed FA-MSVR method outperforms all of the other counterparts.

Comparing the proposed FA-MSVR and the SVR, the FA-MSVR is consistently the best-performing method for ITS forecasting. This ranking attests to the value that is added by simultaneously fitting both the lower and upper bounds of an interval-valued stock index using the MSVR.

When considering the comparison among the three selected interval-valued forecasting methods, we can see that, regardless of the prediction horizon and dataset, the iMLP and VECM achieve better, more accurate forecasts than $Holt^I$, except for the Nikkei 225 index with $h = 5$. Comparing the iMLP and VECM, the performance measures for the iMLP and VECM are mixed.

In all cases, the univariate technique (SVR) invariably has the worst performance. It is conceivable that the reason for the inferiority of the univariate technique for ITS forecasting is that it ignores the possible mutual dependency, i.e., cointegration, between the lower and upper bounds of an interval-valued stock index series.

For each index and prediction horizon, we perform an ANOVA procedure to determine if a statistically significant difference exists among the five methods in the hold-out sample. All of the ANOVA results shown in Table 5 are significant at the 0.05 level, suggesting that there are significant differences among the five methods. To further identify the significant difference between any two methods, the Tukey's HSD test is used to compare all pairwise differences simultaneously at the 0.05 level in the current study. Table 6 shows the results of these multiple comparison tests. For each index and prediction horizon, we rank the methods from 1 (the best) to 5 (the worst).

**\<Insert Tables 5-6 here\>**

Several conclusions can be drawn from Tables 5 and 6. First, when the proposed FA-MSVR method is treated as the testing target, the mean difference between the two adjacent methods is significant at the 0.05 level, indicating that the FA-MSVR performs the best in forecasting interval-valued stock index, under a confidence level of 95%. Exceptions occur when considering the S&P 500 index with $h = 3$ and the Nikkei 225 index with $h = 1$. Second, in three scenarios, the VECM and iMLP yield better results than the Holt[I] with a statistical significance of 95%. Third, the difference between the VECM and iMLP is not significant at the 0.05 level, with two exceptions, where the VECM significantly outperforms the iMLP. Finally, it can be proven that the SVR is the poorest performer at a 95% statistical confidence level in most cases, though there are a few exceptions.

It is important to note that the computational costs of each model for multi-step-ahead prediction are different. From a practical viewpoint, the computational cost is an important and critical issue. Thus, the computational load of each model at each prediction horizon is compared in the present study. Table 7 summarizes the elapsed times for multi-step-ahead forecasting on a hold-out sample for a single replicate. As per the results in Table 7, one can deduce the following observations. Overall, comparing the artificial intelligence models (i.e., FA-MSVR, iMLP, and SVR) with the statistical models (i.e., VECM, Holt[I]), the statistical models are less expensive. The iMLP is computationally much more expensive than the FA-MSVR and SVR for ITS forecasting. When comparing the FA-MSVR and SVR, the FA-MSVR is the winner. The Holt[I] is the least expensive method for ITS forecasting, as its computational requirement is negligible. The elapsed time for ITS forecasting using VECM is also very small.

**<Insert Table 7 here>**

## 5.2 Assessing relative forecast performance with economic criteria in a simple trading strategy

In this section, we assess the economic criteria, using the interval forecasts (one- and multi-step-ahead forecasts) from the different methods in a simple trading strategy, to determine whether the FA-MSVR is a superior forecasting approach for an

investor.[4]

To evaluate the performance of the different methods in terms of economic criteria, we simulate the buy/sell actions according to the simple trading strategy described in Section 4.2 over the period December 2, 2011, to August 10, 2012, for the S&P 500 index; December 30, 2010, to August 10, 2012, for the FTSE 100 index; and August 31, 2009, to August 10, 2012, for the Nikkei 225 index, which are also the out-of-sample forecasting periods for each stock index considered in Section 4.1. According to the work of He, et al. [37], some remarks regarding the actual implementation of this trading strategy are in order. First, if the number of consecutive trading days $k$ chosen is too small, then it can result in substantial noise during training, while some profitable trading opportunities may be forgone if $k$ is too large [37]. Following the works of Cheung, et al. [14] and He, et al. [37], $k$ is experimentally set to 1, 2, and 3. Second, to mimic the transaction cost, we include a one-time 0.1% deduction. Third, the investors are supposed to enter the market at any time during the evaluation period. Fourth, at the end of the evaluation period, if *Step 3* is still in process, then the assets that are not yet sold will not be considered in the profit calculation.

We assess the relative forecast performance in terms of the economic criteria using interval forecasts obtained from the examined methods for three stock indices. Table 8 shows the average return and percentage of trades resulting in positive returns for each scenario. Note that all returns are expressed in annualized terms because every trade has a different asset holding period [37]. Several observations can be drawn from the results in Panel A, which presents the profitability of trading the S&P 500 index derived from the trading signals based on the values of $\hat{X}_{t+h}^{L}$ and $\hat{X}_{t+h}^{U}$ $(h = 1, 3, 5)$ that are obtained from different methods and with $k = 1, 2, \text{and } 3$.

---

[4] As Satchell and Timmermann [40] noted, standard forecasting criteria are not necessarily particularly well suited to assess the economic value of prediction of a nonlinear process.

<center>**<Insert Table 8 here>**</center>

Overall, the five examined methods result in positive returns more often than negative returns by means of the trading strategy for the overwhelming majority of scenarios considered. Indeed, across various parameter combinations and different methods, the percentage of profitable trades is always larger than 50% and could reach a high of 84%. Some exceptions occur when $k = h = 1$ and the SVR forecasts are used or $k = 1, h = 5$ and the Holt[I] forecasts are used; in these scenarios, the trading strategies based on the SVR and Holt[I] forecasts result in profitable trades in only 43.44% and 43.28% of the time, respectively. Another exception occurs when $k = 2, h = 3$ and the SVR forecasts are used; in this scenario, the frequency of profitable trades is roughly on par with that of losses. It should be noted that the highest percentage of profitable trades occurs when $k = 2, h = 3$ and the FA-MSVR forecasts are used; the trading strategy results in profitable trades in 84.44% of cases.

The average annualized returns also suggest that the examined methods perform quite well. The annualized return, which spans from a low of 22.87% to a high of 64.33%, is quite variable between the methods and between the choices for $k$ and $h$. It is worth noting that the smallest average annualized return occurs when $k = 1, h = 1$ and the SVR forecasts are used, while the largest average annualized return occurs when $k = 1, h = 3$ and the FA-MSVR forecasts are used.

When considering the comparison among the five examined methods in terms of the percentage of trades with a positive annualized return, we can see that whatever the choices of $h$ and $k$, the proposed FA-MSVR outperforms all other competitors. The only exception is that the iMLP performs better when $k = 3$ and $h = 1$. Comparing the three selected interval-valued forecasting methods, the ranks of the performance measures for iMLP, VECM, and Holt[I] are mixed. Recall that the SVR invariably has the worst performance in terms of ARV[I] presented in Section 5.1. It is worth noting that the SVR outperforms the Holt[I] when $k = 1$ and $h = 5$، using the economic criteria of the percentage of trades with positive annualized return.

Panels B and C of <span style="color:red">Table 8</span> present the results of using FTSE 100 and Nikkei 225 as the trading index, respectively. Broadly speaking, the results pertaining to the FTSE 100 and Nikkei 225 are quite comparable to those in Panel A. For brevity, we do not repeat the observations, but some overall comments on the performance across all three data sets are made as follows: (1) The profitable trades usually outnumber those incurring a loss. (2) The average or expected annualized returns are always more than 21%. (3) The FA-MSVR performs strikingly better than all other counterparts, despite a few exceptions. Thus, this leads to the fourth comment: (4) The proposed FA-MSVR method can be used as a promising solution for making investor decisions in a financial market.

# 6. Conclusions

Interval forecasting of a stock price index plays an increasingly important role in the financial market, as it could prove to be a potential tool for both private and institutional investors to make profits as well as to avoid risks. In this paper, we introduced a novel hybrid method, a multi-output SVR optimized by a FA (FA-MSVR), for interval forecasting of three globally traded broad market indices (the S&P 500 for the US, the FTSE 100 for the UK, and the Nikkei 225 for Japan). The experimental study was carried out on the basis of various statistical criteria and by assessing the economic values of the predictors. In both cases, we use different forecast horizons (one- and multi-step-ahead) to analyze the robustness of the results. The results obtained show that the proposed FA-MSVR method can statistically outperform some well-established counterparts in terms of the forecast accuracy measure and the accuracy of competing forecasts and, more importantly, can successfully make a profit using a simple trading strategy. These results indicate that the proposed FA-MSVR method is a promising alternative for interval-valued financial time series forecasting problems.

In addition to stock price, the proposed model might be used for other tough interval-valued time series forecasting tasks in a financial market such as an exchange rate, which should be studied in the future. Furthermore, this study challenges the exclusive attention paid to time series forecasting. Econometric models, which reveal the relationship between stock price and selected technical analysis indicators, are of greater value to decision-makers than time series forecasts in financial markets. We will study these issues in future research.

## Acknowledgments

This work was supported by the Natural Science Foundation of China under Project No. 70771042, the Fundamental Research Funds for the Central Universities (2012QN208-HUST), the MOE (Ministry of Education in China) Project of Humanities and Social Science (Project No. 13YJA630002), and a grant from the Modern Information Management Research Center at Huazhong University of Science and Technology (2013WZ005 2012WJD002).

## Appendix A: FA-MSVR vs. PSO-MSVR vs. GA-MSVR

In this Appendix A, we have compared with particle swarm optimization (PSO) and genetic algorithms (GA) along with thorough discussion on the basis of the prediction accuracy, economic criteria, and computational time. However, it should be noted that the aim of this study is to originally investigate the possibility of forecasting the interval-valued stock price index series over short and long horizons using multi-output support vector regression, and conduct a large scale comparative study with the selected well-established interval-valued forecasting methods (i.e., iMLP, Holt[I], and VECM). Concerning the parameter selection of MSVR, the superiority of firefly algorithm (FA) against GA and PSO in existing studies [29-31] motivates us to use the FA for parameters selection of MSVR. Although, the comparison among FA, PSO, and GA is of great importance, it could make the article a little bit redundant if we add more details on comparison among FA, PSO, and GA into the main text of the revised manuscript. Thus the results are given in this Appendix A.

Tables 9-10 show the comparisons of performance of FA-MSVR, PSO-MSVR, and GA-MSVR in terms of statistical and economic criteria, respectively. The results in Tables 9-10 lead to the following conclusions. When considering the prediction accuracy, The FA-MSVR and PSO-MSVR seem to produce forecasts which are more accurate than those of the GA-MSVR (though only marginally). However, the difference between the FA-MSVR and PSO-MSVR in prediction accuracy is not clear. Concerning the economic criteria, overall, the FA-MSVR, PSO-MSVR, and GA-MSVR are almost a tie.

**<Insert Tables 9-10 here>**

Table 11 summarizes the elapsed times of FA-MSVR, PSO-MSVR, and GA-MSVR for multi-step-ahead forecasting on hold-out sample for a single replicate. According to the obtained results, the GA-MSVR is computationally much more expensive than the FA-MSVR and PSO-MSVR. The FA-MSVR is the least expensive method in this case.

**<Insert Table 11 here>**

We have listed the results of the required comparison, but the experimental findings cannot be conclusive with mixing results in terms of prediction accuracy and economic criteria. The only conclusion we can draw is that the FA-MSVR is the least expensive method in this case. Thus, the firefly algorithm is adopted here for parameter selection of MSVR.

# Caption page

Table 1: Interval-valued variables

Table 2 Period and length of the interval-valued stock price index time series processed

Table 3 Effects of population size for FA-MSVR on prediction accuracy

Table 4 Cointegration test results for interval-valued S&P 500 index series

Table 5 ANOVA test results for hold-out sample

Table 6 Multiple comparison results with ranked methods for hold-out sample

Table 7 Required time of examined models for each prediction horizon

Table 8 Performance comparison of different methods using a trading strategy for hold-out sample

Table 9 Performance comparison of three heuristic algorithm-based MSVR methods in terms of ARV[I]

Table 10 Performance comparison of three heuristic algorithm-based MSVR methods using a trading strategy for hold-out sample

Table 11 Required time of three heuristic algorithm-based MSVR methods for each prediction horizon

Fig. 1: Interval structure

Fig. 2: (a) 10-minutes S&P 500 index; (b) Corresponding daily S&P 500 index intervals.

Fig. 3: Pseudocode for the FA-MSVR method

Fig. 4. Interval S&P 500 index (the estimation sample is the earlier section and the hold-out sample is the later section)

Fig. 5: Pseudocode for the calculation of the relevance criterion

Fig. 6: Performance comparison of different methods in terms of ARV[I] on interval-valued S&P500 index series

Fig. 7: Performance comparison of different methods in terms of ARV[I] on interval-valued FTSE 100 index series

Fig. 8: Performance comparison of different methods in terms of ARV[I] on interval-valued Nikkei 225 index series.

# Figures

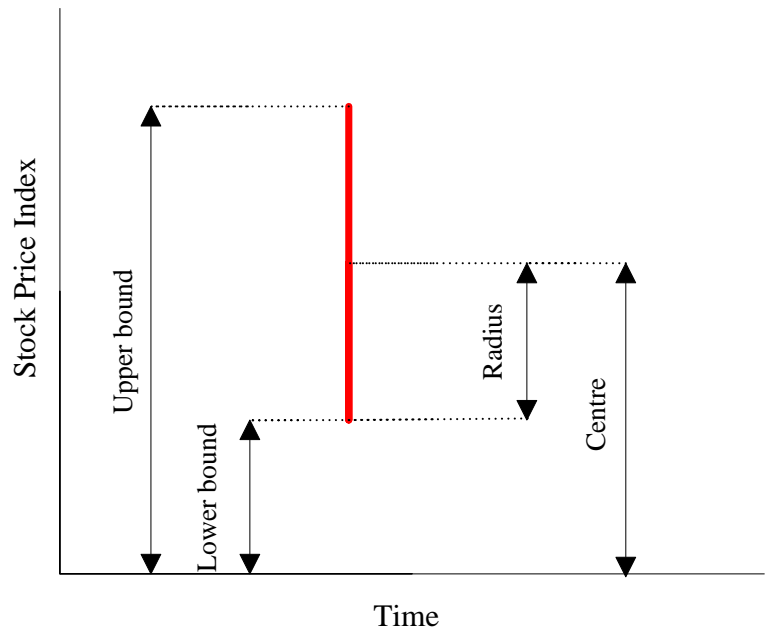

Fig. 1. Interval structure

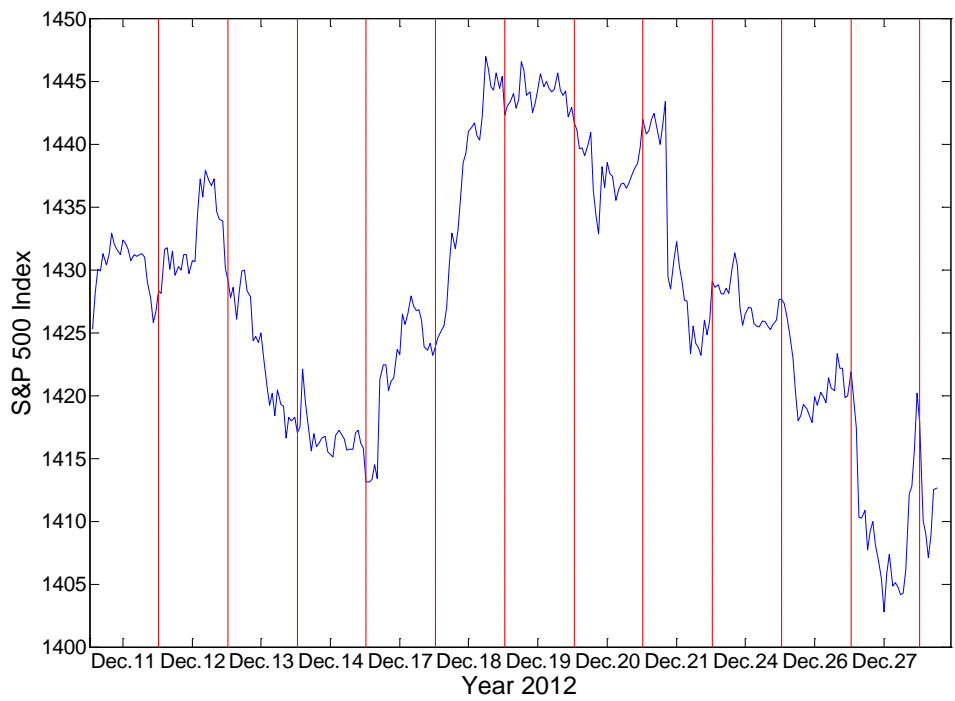

(a)

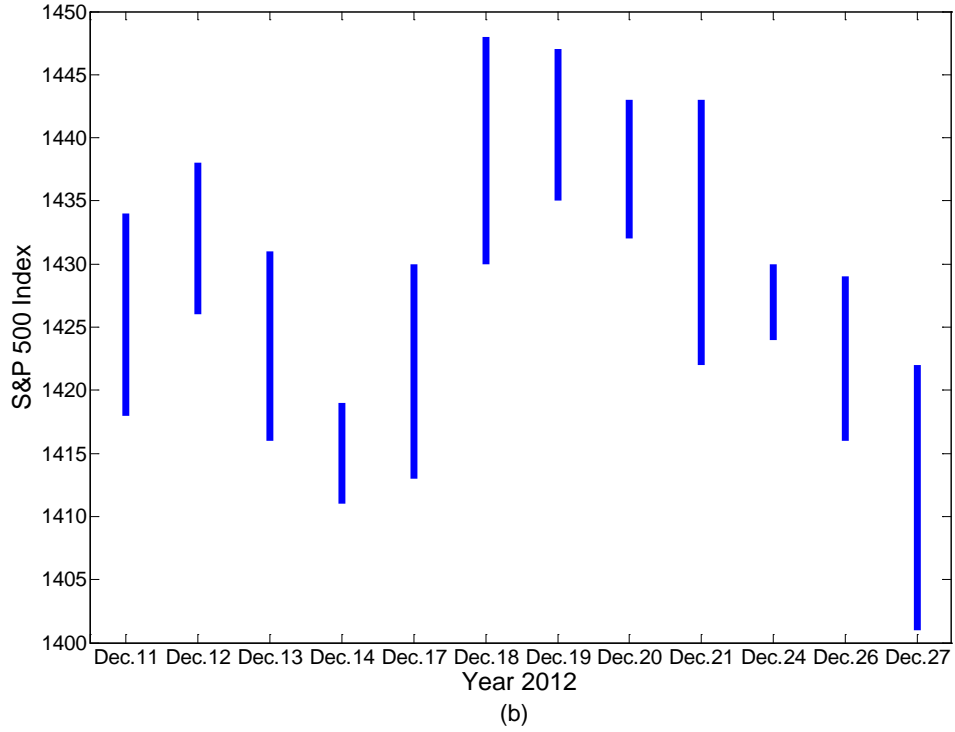

Fig. 2. (a) 10-minutes S&P 500 index; (b) Corresponding daily S&P 500 index intervals.

---

**Algorithm 1 FA-MSVR method**

---

**Define** fitness function $f(x)$ according to Eq. (6), $x = (\log_2 C, \log_2 \sigma, \log_2 \varepsilon) \in \mathbb{R}^3$

**Initialize** a population of firefly $x_i$ $(i = 1, 2, \ldots, n)$.

**Define** light absorption coefficient $\gamma$.

**While** $(t < \text{maxGeneration})$

    **for** $i = 1 : n$ all $n$ fireflies

        **for** $j = 1 : i$ all $n$ fireflies

        Light intensity $I_i^t$ at $x_i^t$ is determined by $f(x_i^t)$

        **if** $(I_j^t > I_i^t)$

        Move firefly $x_i^t$ towards $x_j^t$ in all three dimensions according to Eq. (5)

        **End if**

        Attractiveness varies with distance $r$ via $\exp\left[-\gamma r^2\right]$

        Evaluate new solutions and update light intensity

        **End for** $j$

    **End for** $i$

**End while**

---

Fig. 3. Pseudocode for the FA-MSVR method

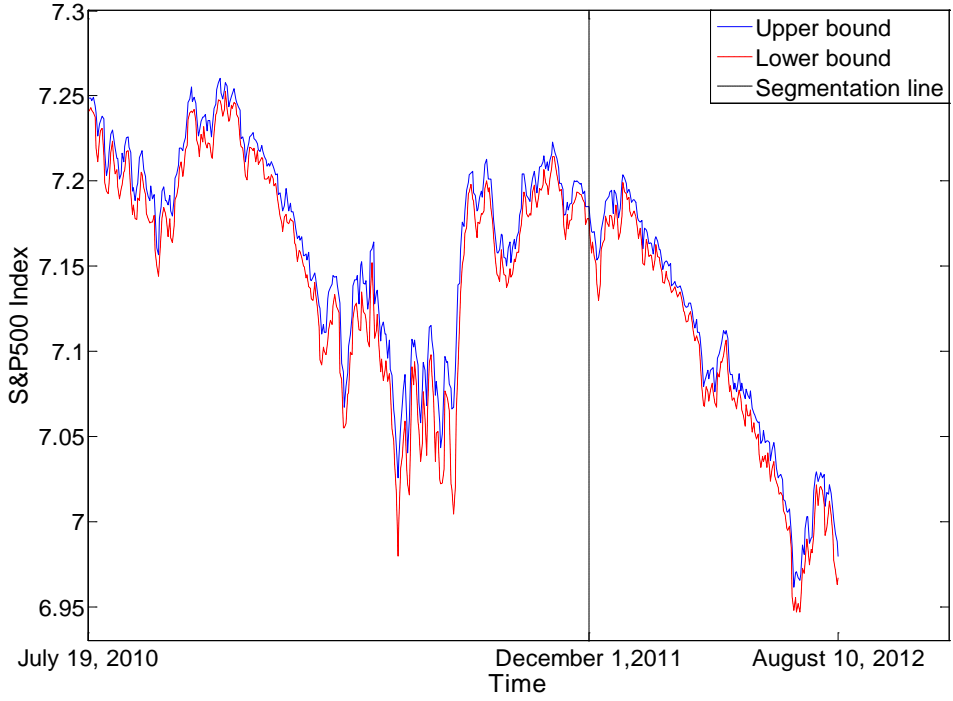

Fig. 4. Interval S&P 500 index (the estimation sample is the earlier section and the hold-out sample is the later section)

---

**Algorithm 2** Calculation of the relevance criterion

**Given** a dataset $D = \left\{ \left( \mathbf{x}_i, \mathbf{y}_i \right) \in \left( \mathbb{R}^{2d} \times \mathbb{R}^2 \right) \right\}_{i=d}^{N}$, where $\mathbf{x}_i = \left[ X_i^L, X_i^U, \ldots, X_{i-d+1}^L, X_{i-d+1}^U \right]^T$ and $\mathbf{y}_i = \left[ X_{i+1}^L, X_{i+1}^U \right]^T$

**Given** a set of variable $V$ of size $m \left( m \leq d \right)$, with $V \subset \left\{ X_{.1}^L, X_{.1}^U, \ldots, X_{.d}^L, X_{.d}^U \right\}$.

**Given** a metric on the space $\mathbb{R}^{2d}$

**Divide** the $N - d + 1$ input-output pairs in two parts $D_1$ and $D_2$

**For** each point $\left( \mathbf{x}_i, \mathbf{y}_i \right)$ in $D_1$

- Find the nearest neighbor, say $\hat{\mathbf{x}}_i$, of $\mathbf{x}_i$ in $D_2$ according to the metric and the set $V$.

- Calculate $error_{\mathbf{x}_i} = \frac{1}{2} \left( \left( \hat{X}_{i+1}^L - X_{i+1}^L \right)^2 + \left( \hat{X}_{i+1}^U - X_{i+1}^U \right)^2 \right)$ where $\left[ X_{i+1}^L, X_{i+1}^U \right]$ is the output of $\mathbf{x}_i$ and $\left[ \hat{X}_{i+1}^L, \hat{X}_{i+1}^U \right]$ is the output of $\hat{\mathbf{x}}_i$.

**End**

**For** each point $\left( \mathbf{x}_i, \mathbf{y}_i \right)$ in $D_2$

- Find the nearest neighbor, say $\hat{\mathbf{x}}_i$, of $\mathbf{x}_i$ in $D_1$ according to the metric and the set $V$.

- Calculate $error_{\mathbf{x}_i} = \frac{1}{2} \left( \left( \hat{X}_{i+1}^L - X_{i+1}^L \right)^2 + \left( \hat{X}_{i+1}^U - X_{i+1}^U \right)^2 \right)$ where $\left[ X_{i+1}^L, X_{i+1}^U \right]$ is the output of $\mathbf{x}_i$ and

$\left[ \bar{X}_{i+1}^L, \bar{X}_{i+1}^U \right]$ is the output of $\hat{\mathbf{x}}_i$.

**Calculate** $E(V) = \dfrac{1}{N-d+1} \sum\limits_{i=d}^{N} error_{\hat{\mathbf{x}}_i}$ which is the statistical measure of the relevance of the set of variables $V$.

Fig. 5. Pseudocode for the calculation of the relevance criterion

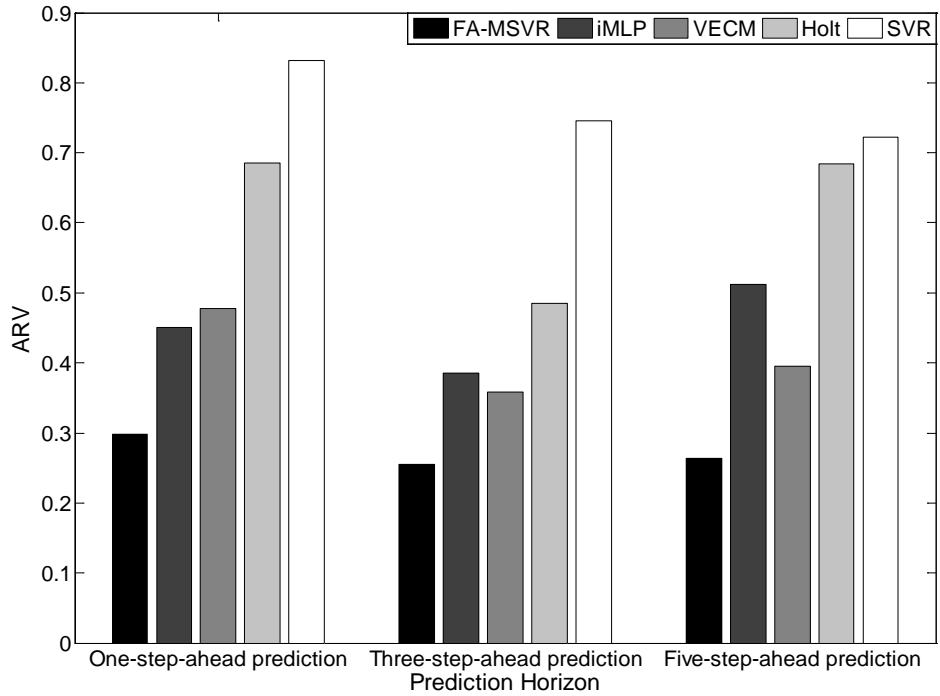

Fig. 6. Performance comparison of different methods in terms of ARV[1] on interval-valued S&P500 index series

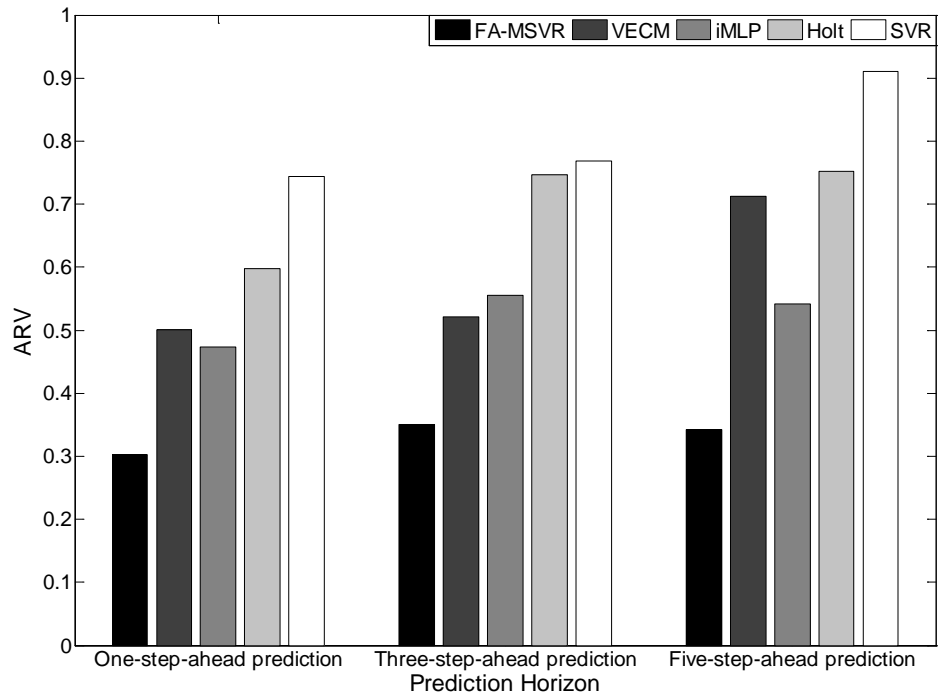

Fig. 7. Performance comparison of different methods in terms of ARV[I] on interval-valued FTSE 100 index series

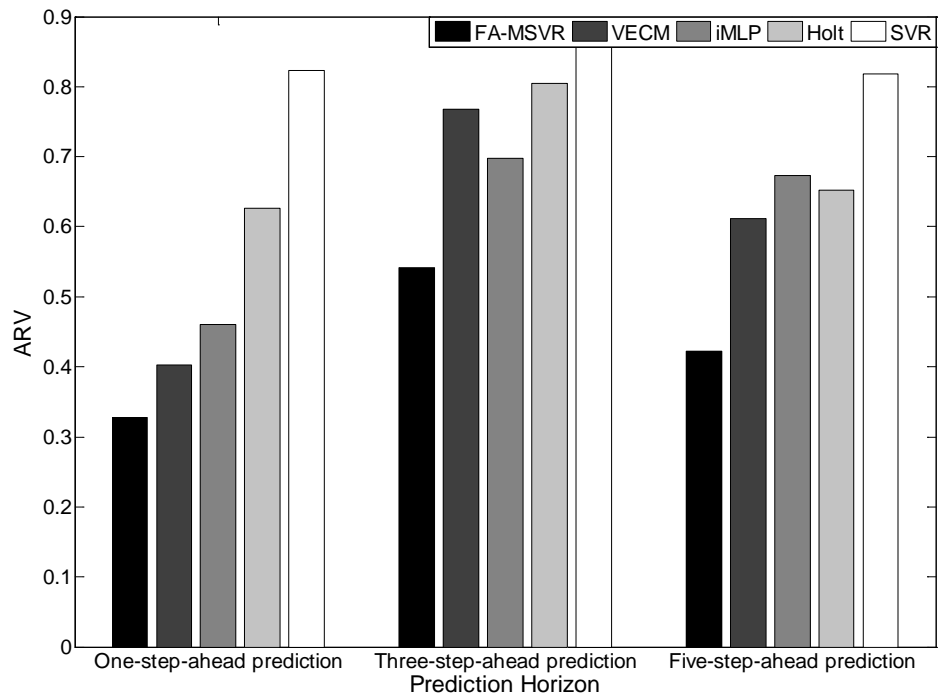

Fig. 8. Performance comparison of different methods in terms of ARV[I] on interval-valued Nikkei 225 index series.

# Tables

Table 1 Interval-valued variables

| Year 2012 | S&P 500 index |
| | [Lower, Upper] |
| --- | --- |
| Dec. 11 | [1418.55, 1434.27] |
| Dec. 12 | [1426.76, 1438.59] |
| Dec. 13 | [1416, 1431.36] |
| Dec. 14 | [1411.88, 1419.45] |
| Dec. 17 | [1413.54, 1430.67] |
| … | … |
| … | … |

Table 2 Period and length of the interval-valued stock price index time series processed

| ITS | Period | Sample size |
| --- | --- | --- |
| S&P 500 | July 19, 2010 to August 10, 2012 | 523 |
| FTSE 100 | October 15, 2007 to August 10, 2012 | 1218 |
| Nikkei 225 | January 8, 2004 to August 10, 2012 | 2165 |

Table 3 Effects of population size for FA-MSVR on prediction accuracy

| Prediction horizon | Population size | | | | |
| | 10 | 15 | 20 | 25 | 30 |
| --- | --- | --- | --- | --- | --- |
| *Panel A:* S&P 500 index | | | | | |
| $h=1$ | 0.294 | 0.301 | 0.299 | 0.294 | 0.299 |
| $h=3$ | 0.261 | 0.256 | 0.255 | 0.247 | 0.252 |
| $h=5$ | 0.274 | 0.258 | 0.263 | 0.249 | 0.273 |
| | | | | | |
| *Panel B:* FTSE 100 index | | | | | |
| $h=1$ | 0.309 | 0.308 | 0.301 | 0.296 | 0.298 |
| $h=3$ | 0.345 | 0.353 | 0.351 | 0.349 | 0.343 |
| $h=5$ | 0.350 | 0.352 | 0.340 | 0.326 | 0.341 |
| | | | | | |
| *Panel C:* Nikkei 225 index | | | | | |
| $h=1$ | 0.317 | 0.324 | 0.324 | 0.332 | 0.322 |
| $h=3$ | 0.534 | 0.539 | 0.541 | 0.550 | 0.536 |
| $h=5$ | 0.429 | 0.444 | 0.432 | 0.431 | 0.441 |

Table 4 Cointegration test results for interval-valued S&P 500 index series

|      | EIGENV | TRACE   | U      | L        | LAG |
|------|--------|---------|--------|----------|-----|
| r=1  | 0.007  | 3.736   |        |          | 5   |
| r=0  | 0.045* | 27.677* |        |          | 5   |
| Q(6) |        |         | 0.9772 | 1.274    |     |
| Q(12)|        |         | 3.4506 | 10.584   |     |
|      | C      |         | 1      | -0.97097 |     |

*Notes*: Eigenvalue and trace statistics are given under the columns 'EIGENV' and 'TRACE.' 'r=0' corresponds to the null hypothesis of no cointegration and 'r=1' corresponds to the hypothesis of one cointegration vector. * denotes rejection of the hypothesis at the 0.05 level. The no-cointegration null is rejected and the hypothesis of one-cointegration vector is not rejected. 'U' and 'L' identify the Q-statistics associated with the daily upper and lower bound series equations. All the Q-statistics are insignificant. The rows labeled 'C' give cointegrating vectors with the coefficient of the upper bound series normalized to one. 'LAG' the lag parameters used to conduct the test.

Table 5 ANOVA test results for hold-out sample

| Prediction horizon | ANOVA test | |
|---|---|---|
| | Statistics $F$ | $p$ -value |
| *Panel A:* S&P 500 index | | |
| $h=1$ | 22.548 | 0.000* |
| $h=3$ | 19.654 | 0.000* |
| $h=5$ | 31.254 | 0.000* |
| | | |
| *Panel B:* FTSE 100 index | | |
| $h=1$ | 14.121 | 0.000* |
| $h=3$ | 8.597 | 0.000* |
| $h=5$ | 26.854 | 0.000* |
| | | |
| *Panel C:* Nikkei 225 index | | |
| $h=1$ | 32.484 | 0.000* |
| $h=3$ | 8.497 | 0.001* |
| $h=5$ | 13.524 | 0.000* |

*Notes*: *indicates the mean difference among the five methods is significant at the 0.05 level.

Table 6 Multiple comparison results with ranked methods for hold-out sample

| Prediction Horizon | Rank of methods | | | | | | | | |
|---|---|---|---|---|---|---|---|---|---|
| | 1 | | 2 | | 3 | | 4 | | 5 |
| *Panel A:* S&P 500 index | | | | | | | | | |
| $h=1$ | FA-MSVR | <* | iMLP | < | VECM | <* | Holt[I] | <* | SVR |
| $h=3$ | FA-MSVR | < | VECM | < | iMLP | < | Holt[I] | <* | SVR |
| $h=5$ | FA-MSVR | <* | VECM | <* | iMLP | < | Holt[I] | < | SVR |
| *Panel B:* FTSE 100 index | | | | | | | | | |
| $h=1$ | FA-MSVR | <* | VECM | < | iMLP | < | Holt[I] | <* | SVR |
| $h=3$ | FA-MSVR | <* | iMLP | < | VECM | <* | Holt[I] | < | SVR |
| $h=5$ | FA-MSVR | <* | VECM | <* | iMLP | < | Holt[I] | < | SVR |
| *Panel C:* Nikkei 225 index | | | | | | | | | |
| $h=1$ | FA-MSVR | < | iMLP | < | VECM | <* | Holt[I] | <* | SVR |
| $h=3$ | FA-MSVR | <* | VECM | < | iMLP | < | Holt[I] | < | SVR |
| $h=5$ | FA-MSVR | <* | iMLP | < | VECM | < | Holt[I] | <* | SVR |

*Notes*: *indicates the mean difference between the two adjacent methods is significant at the 0.05 level.

Table 7 Required time of examined models for each prediction horizon

| Prediction Horizon | Elapsed time (s) | | | | |
|---|---|---|---|---|---|
| | FA-MSVR | iMLP | VECM | Holt[I] | SVR |
| *Panel A:* S&P 500 index | | | | | |
| $h=1$ | 18.425 | 34.121 | 0.125 | 0.054 | 23.541 |
| $h=3$ | 18.435 | 34.131 | 0.126 | 0.053 | 23.551 |
| $h=5$ | 18.444 | 34.140 | 0.126 | 0.060 | 23.560 |
| *Panel B:* FTSE 100 index | | | | | |
| $h=1$ | 18.449 | 34.145 | 0.129 | 0.058 | 23.565 |
| $h=3$ | 18.450 | 34.146 | 0.130 | 0.059 | 23.566 |
| $h=5$ | 18.455 | 34.151 | 0.132 | 0.061 | 23.571 |
| *Panel C:* Nikkei 225 index | | | | | |
| $h=1$ | 18.491 | 34.125 | 0.120 | 0.055 | 23.609 |
| $h=3$ | 18.511 | 34.205 | 0.124 | 0.058 | 23.568 |
| $h=5$ | 18.515 | 34.143 | 0.123 | 0.061 | 23.570 |

Table 8 Performance comparison of different methods using a trading strategy for hold-out sample

| Prediction horizon | Method | $k=1$ | | $k=2$ | | $k=3$ | |
|---|---|---|---|---|---|---|---|
| | | AVERAGE | POSITIVE | AVERAGE | POSITIVE | AVERAGE | POSITIVE |
| *Panel A:* S&P 500 index | | | | | | | |
| | FA-MSVR | 38.58% | 63.37% | 45.92% | 71.41% | 44.01% | 65.58% |
| | iMLP | 36.33% | 57.00% | 47.28% | 66.01% | 40.44% | 70.70% |
| $h=1$ | VECM | 24.28% | 61.92% | 38.43% | 68.31% | 34.44% | 64.44% |
| | Holt[t] | 27.35% | 52.35% | 23.03% | 58.74% | 39.57% | 58.32% |
| | SVR | 22.87% | 43.44% | 27.85% | 57.30% | 26.00% | 56.00% |
| | | | | | | | |
| | FA-MSVR | 64.33% | 62.01% | 58.90% | 84.44% | 52.30% | 80.81% |
| | iMLP | 43.00% | 59.58% | 32.01% | 77.70% | 48.28% | 75.37% |
| $h=3$ | VECM | 46.01% | 60.45% | 43.01% | 80.67% | 35.01% | 62.74% |
| | Holt[t] | 38.90% | 58.65% | 28.70% | 52.18% | 40.67% | 65.97% |
| | SVR | 30.01% | 54.43% | 30.44% | 50.18% | 25.01% | 54.44% |
| | | | | | | | |
| | FA-MSVR | 49.18% | 70.20% | 51.04% | 79.12% | 46.92% | 78.30% |
| | iMLP | 39.39% | 72.00% | 49.12% | 67.55% | 37.81% | 74.24% |
| $h=5$ | VECM | 42.74% | 55.02% | 37.97% | 63.60% | 49.79% | 67.14% |
| | Holt[t] | 36.73% | 43.28% | 31.97% | 58.10% | 34.22% | 52.77% |
| | SVR | 27.11% | 51.99% | 32.50% | 55.32% | 23.96% | 51.52% |
| *Panel B:* FTSE 100 index | | | | | | | |
| | FA-MSVR | 42.31% | 68.68% | 38.70% | 73.70% | 51.40% | 65.40% |
| $h=1$ | iMLP | 38.40% | 64.44% | 35.01% | 68.61% | 41.90% | 67.44% |
| | VECM | 29.68% | 52.61% | 28.70% | 62.34% | 37.37% | 55.01% |

|  |  |  |  |  |  |  |
|---|---|---|---|---|---|---|
| | Holt[t] | 31.41% | 48.01% | 30.74% | 52.64% | 22.00% | 57.21% |
| | SVR | 25.01% | 53.74% | 24.34% | 58.61% | 28.58% | 51.43% |
| | FA-MSVR | 50.01% | 71.70% | 43.01% | 65.00% | 54.01% | 73.01% |
| | iMLP | 50.37% | 65.44% | 41.43% | 71.40% | 37.64% | 67.00% |
| $h=3$ | VECM | 42.61% | 68.28% | 32.34% | 63.00% | 41.43% | 68.63% |
| | Holt[t] | 35.37% | 56.94% | 28.62% | 42.71% | 35.00% | 53.00% |
| | SVR | 38.80% | 42.01% | 37.01% | 54.67% | 27.62% | 57.01% |
| | FA-MSVR | 45.33% | 68.10% | 39.15% | 72.12% | 42.08% | 77.10% |
| | iMLP | 43.94% | 64.00% | 25.49% | 61.73% | 27.39% | 68.00% |
| $h=5$ | VECM | 51.10% | 68.79% | 39.02% | 63.81% | 39.60% | 71.11% |
| | Holt[t] | 33.09% | 51.02% | 28.36% | 52.85% | 30.75% | 60.57% |
| | SVR | 34.82% | 44.11% | 25.13% | 50.93% | 25.10% | 59.50% |

*Panel C:* Nikkei 225 index

|  |  |  |  |  |  |  |  |
|---|---|---|---|---|---|---|---|
| | FA-MSVR | 42.70% | 71.01% | 34.43% | 62.68% | 39.38% | 67.31% |
| | iMLP | 38.61% | 66.01% | 40.34% | 65.01% | 31.57% | 58.61% |
| $h=1$ | VECM | 31.68% | 57.64% | 32.34% | 52.01% | 24.01% | 61.43% |
| | Holt[t] | 36.33% | 61.01% | 25.03% | 50.16% | 25.34% | 53.44% |
| | SVR | 28.28% | 47.01% | 26.34% | 54.01% | 21.20% | 51.90% |
| | FA-MSVR | 51.44% | 66.91% | 43.01% | 78.70% | 60.29% | 71.62% |
| | iMLP | 43.37% | 70.67% | 37.61% | 72.70% | 47.67% | 66.01% |
| $h=3$ | VECM | 47.01% | 62.34% | 36.00% | 69.01% | 28.62% | 61.24% |
| | Holt[t] | 32.01% | 57.64% | 28.61% | 57.29% | 35.01% | 46.01% |

| | | | | | | |
|---|---|---|---|---|---|---|
| | SVR | 23.74% | 54.01% | 24.95% | 52.34% | 28.61% | 52.59% |
| | FA-MSVR | 43.45% | 69.49% | 44.68% | 80.54% | 35.52% | 72.53% |
| $h=5$ | iMLP | 45.50% | 63.88% | 39.37% | 67.11% | 42.06% | 73.27% |
| | VECM | 36.45% | 56.95% | 30.38% | 70.83% | 32.39% | 62.70% |
| | Holt[I] | 31.12% | 50.72% | 38.98% | 54.71% | 31.04% | 60.46% |
| | SVR | 29.92% | 43.43% | 22.63% | 53.62% | 27.98% | 55.28% |

*Notes*: Average annualized returns and percentage of trades with positive annualized returns are given under the columns 'AVERAGE' and 'POSITIVE'. All returns in Table 6 are expressed in annualized terms. Let $X_t^C$ and $X_{t+i}^C$ $(i \geq 1)$ be, respectively, the closing index on the buying and selling day. The actual percentage return of a given trade, net of transaction cost, is defined by $R = \left( X_{t+i}^C - X_t^C \big/ X_t^C \right) \cdot 100\% - 0.1\%$, while the corresponding annualized return is $AR = \left( R/i \right) \cdot 365$.

Table 9 Performance comparison of three heuristic algorithm-based MSVR methods in terms of ARV[1]

| Prediction horizon | Method | | |
| --- | --- | --- | --- |
| | FA-MSVR | PSO-MSVR | GA-MSVR |
| *Panel A:* S&P 500 index | | | |
| $h = 1$ | 0.299 | 0.312 | 0.287 |
| $h = 3$ | 0.255 | 0.267 | 0.255 |
| $h = 5$ | 0.263 | 0.271 | 0.274 |
| | | | |
| *Panel B:* FTSE 100 index | | | |
| $h = 1$ | 0.301 | 0.302 | 0.283 |
| $h = 3$ | 0.351 | 0.335 | 0.365 |
| $h = 5$ | 0.340 | 0.319 | 0.337 |
| | | | |
| *Panel C:* Nikkei 225 index | | | |
| $h = 1$ | 0.324 | 0.309 | 0.341 |
| $h = 3$ | 0.541 | 0.548 | 0.549 |
| $h = 5$ | 0.432 | 0.422 | 0.425 |

Table 10 Performance comparison of three heuristic algorithm-based MSVR methods using a trading strategy for hold-out sample

| Prediction horizon | Method | $k=1$ | | $k=2$ | | $k=3$ | |
|---|---|---|---|---|---|---|---|
| | | AVERAGE | POSITIVE | AVERAGE | POSITIVE | AVERAGE | POSITIVE |
| *Panel A:* S&P 500 index | | | | | | | |
| | FA-MSVR | 38.58% | 63.37% | 45.92% | 71.41% | 44.01% | 65.58% |
| $h=1$ | PSO-MSVR | 39.27% | 64.15% | 44.19% | 70.43% | 42.91% | 66.25% |
| | GA-MSVR | 39.96% | 62.75% | 47.04% | 72.11% | 42.04% | 65.99% |
| | FA-MSVR | 64.33% | 62.01% | 58.90% | 84.44% | 52.30% | 80.81% |
| $h=3$ | PSO-MSVR | 63.88% | 63.67% | 56.90% | 84.29% | 52.00% | 80.65% |
| | GA-MSVR | 65.41% | 61.30% | 60.04% | 84.33% | 50.44% | 79.51% |
| | FA-MSVR | 49.18% | 71.20% | 51.04% | 79.12% | 46.92% | 78.30% |
| $h=5$ | PSO-MSVR | 48.30% | 71.00% | 51.09% | 78.95% | 47.23% | 78.56% |
| | GA-MSVR | 48.76% | 71.06% | 50.07% | 80.09% | 46.25% | 77.51% |
| *Panel B:* FTSE 100 index | | | | | | | |
| | FA-MSVR | 42.31% | 68.68% | 38.70% | 73.70% | 51.40% | 65.40% |
| $h=1$ | PSO-MSVR | 41.46% | 67.04% | 39.00% | 74.43% | 51.59% | 65.10% |
| | GA-MSVR | 42.89% | 69.27% | 39.42% | 74.24% | 53.18% | 64.24% |
| | FA-MSVR | 50.01% | 71.70% | 43.01% | 65.00% | 54.01% | 73.01% |
| $h=3$ | PSO-MSVR | 50.85% | 70.64% | 41.49% | 65.43% | 53.81% | 72.84% |
| | GA-MSVR | 50.66% | 72.78% | 42.41% | 65.65% | 53.67% | 74.38% |
| $h=5$ | FA-MSVR | 45.33% | 68.10% | 39.15% | 72.12% | 42.08% | 77.10% |

|  |  | AVERAGE | POSITIVE | AVERAGE | POSITIVE | AVERAGE | POSITIVE |
|---|---|---|---|---|---|---|---|
|  | PSO-MSVR | 45.07% | 67.50% | 39.13% | 71.80% | 42.98% | 77.94% |
|  | GA-MSVR | 44.44% | 68.58% | 38.69% | 71.97% | 42.18% | 77.99% |
| *Panel C*: Nikkei 225 index | | | | | | | |
|  | FA-MSVR | 42.70% | 71.01% | 34.43% | 62.68% | 39.38% | 67.31% |
| $h=1$ | PSO-MSVR | 44.67% | 69.88% | 32.85% | 61.12% | 37.63% | 66.93% |
|  | GA-MSVR | 42.49% | 70.47% | 35.48% | 63.19% | 40.47% | 69.04% |
|  | FA-MSVR | 51.44% | 66.91% | 43.01% | 78.70% | 60.29% | 71.62% |
| $h=3$ | PSO-MSVR | 53.33% | 65.68% | 41.57% | 79.49% | 58.67% | 71.72% |
|  | GA-MSVR | 51.56% | 68.35% | 42.95% | 78.27% | 60.98% | 72.59% |
|  | FA-MSVR | 43.45% | 69.49% | 44.68% | 80.54% | 35.52% | 72.53% |
| $h=5$ | PSO-MSVR | 43.29% | 70.46% | 44.28% | 80.94% | 35.85% | 72.61% |
|  | GA-MSVR | 43.85% | 69.82% | 44.04% | 79.80% | 36.52% | 71.87% |

*Notes*: Average annualized returns and percentage of trades with positive annualized returns are given under the columns 'AVERAGE' and 'POSITIVE'. All returns in Table 6 are expressed in annualized terms. Let $X_t^C$ and $X_{t+i}^C \left( i \geq 1 \right)$ be, respectively, the closing index on the buying and selling day. The actual percentage return of a given trade, net of transaction cost, is defined by $R = \left( X_{t+i}^C - X_t^C \big/ X_t^C \right) \cdot 100\% - 0.1\%$, while the corresponding annualized return is $AR = \left( R/i \right) \cdot 365$.

Table 11 Required time of three heuristic algorithm-based MSVR methods for each prediction horizon

| Prediction horizon | Elapsed time (s) | | |
|---|---|---|---|
| | FA-MSVR | PSO-MSVR | GA-MSVR |
| *Panel A:* S&P 500 index | | | |
| $h=1$ | 18.425 | 24.657 | 33.880 |
| $h=3$ | 18.435 | 24.628 | 33.981 |
| $h=5$ | 18.444 | 24.683 | 34.016 |
| | | | |
| *Panel B:* FTSE 100 index | | | |
| $h=1$ | 18.449 | 24.638 | 33.921 |
| $h=3$ | 18.450 | 24.677 | 33.977 |
| $h=5$ | 18.455 | 24.645 | 33.945 |
| | | | |
| *Panel C:* Nikkei 225 index | | | |
| $h=1$ | 18.491 | 24.722 | 34.044 |
| $h=3$ | 18.511 | 24.737 | 34.054 |
| $h=5$ | 18.515 | 24.703 | 33.962 |